\documentstyle[prb,twocolumn,aps,epsf]{revtex}
\begin{document}
\twocolumn[
\hsize\textwidth\columnwidth\hsize\csname@twocolumnfalse\endcsname

\title{Collective charge density fluctuations in superconducting
layered systems with bilayer unit cells}
\author{E. H.\ Hwang and S. \ Das Sarma}
\address{Center for Superconductivity Research,Department of Physics \\
	 University of Maryland, College Park, Maryland  20742-4111 }
\date{\today}
\maketitle

\begin{abstract}

Collective modes of  bilayered superconducting
superlattices (e.g., YBCO) are 
investigated within the conserving gauge-invariant ladder diagram
approximation including both the nearest interlayer single
electron tunneling and the Josephson-type Cooper pair tunneling.
By calculating the density-density response function including
Coulomb and pairing interactions, we examine the
two collective mode branches corresponding to 
the in-phase and out-of-phase charge fluctuations between the two
layers in the 
unit cell. The out-of-phase collective mode develops 
a long wavelength plasmon gap whose magnitude depends on the 
tunneling strength with the mode dispersions being
insensitive to the specific tunneling mechanism (i.e., single electron
or Josephson).
We also show that in the presence of tunneling the oscillator
strength of the  out-of-phase mode overwhelms that of the in-phase-mode
at $k_{\|} = 0$ and  finite $k_z$, where $k_z$ and $k_{\|}$ are
respectively the mode wave vectors perpendicular and along the
layer. We discuss the possible experimental observability of the
phase fluctuation modes in the context of our theoretical results for the
mode dispersion and spectral weight. \\

PACS Number : 74.20.-z; 74.80.Dm; 71.45.Gm; 74.25.Gz

\end{abstract}
\vspace{0.5in}
]


\section{introduction}

In contrast to bulk isotropic superconductors, where longitudinal
collective modes (i.e., plasmons) associated with the density fluctuation
response is virtually of no particular interest or significance in the
context of superconducting properties, there has been substantial
recent theoretical interest in the longitudinal collective mode
spectra of layered high-$T_c$ superconductors
\cite{fertig,hwang,cote,wu,for,das}. This interest arises 
primarily from the highly anisotropic two dimensional layered
structure of these materials which, in principle, allow for sub-gap
plasmon modes residing inside the superconducting gap in the low wave
vector regime. This gives rise to interesting collective mode behavior
\cite{fertig,hwang,cote,wu,for,das,kake,kado,dunm,buis} 
in layered anisotropic superconductors which have no analogs in bulk
isotropic superconductors. In this paper we consider the effect of
having multilayer complex unit cells, as existing in YBCO and BISCO
high-$T_c$ superconductor materials, on the longitudinal electronic
collective mode spectrum. We find a number of collective modes arising
from the complex unit cell structure, and comment on their possible
experimental relevance. One of our goals is to critically assess
whether observable electronic collective mode behavior could shed some
light on the interesting and unusual mechanism producing high $T_c$
superconductivity in these materials. The other goal is to predict
novel collective mode behavior peculiar to layered superconductors
with no analogs in bulk systems.

The collective mode spectrum is characterized by the energy dispersion
($\hbar = 1$ throughout this paper)
relation $\omega \equiv \omega(k_{\|},k_z)$, which we calculate in
this paper, where $k_{\|} \equiv |{\bf k}_{\|}|$ is the two dimensional wave
vector in the so-called {\it a-b} plane (along the layer) and $k_z$ is the
wave vector along the $c$-axis, then $k_z = |{\bf k}| \cos\theta$,
$k_{\|} = |{\bf k}|
\sin \theta$.   Because of the strong {\it a-b} plane versus {\it
c}-axis anisotropy in these materials, the dependence of the collective
mode frequency on $k_{\|}$ and $k_z$ is very different. [We ignore any
anisotropy, which is invariably rather small, in the {\it a-b} plane
and assume intralayer planar isotropy, i.e., $\omega({\bf k}_{\|},k_z) \equiv
\omega(k_{\|},k_z)$.] The structural model we employ considers the layered
superconductor to be a one dimensional superlattice along the $z$
direction ($c$-axis) composed of a periodic system of bilayer unit
cells with an intracell layer separation of $c$ and a superlattice
period of $d$ ($>c$). The two active layers separated by a distance
$c$ within each unit cell are taken to
be identical and are assumed to be planar two dimensional electron gas
(2D EG) systems of charge density $n_s$ per unit area and zero layer
thickness each. In most of our calculations presented in this paper
the intercell electron hopping (or tunneling) between neighboring unit
cells (separated by a distance $d$) 
is neglected (i.e., we neglect any superlattice band width along
the $z$ direction), but we critically examine the effect of intracell
electron hopping between the two layers within each unit cell on the
collective mode dispersion. We comment upon the effect of a finite
{\it inter}cell hopping in the conclusion of this article. We include
in our theory the long range (intracell and intercell) Coulomb
interaction among all the layers. This long range Coulomb interaction,
which couples all the layers, is of great importance in determining
the collective mode spectrum. We also include in our theory of
collective mode dispersion the effect of the superconducting pairing
interaction, assumed in our model to be a short-range in-plane
attractive interaction of the BCS-Fermi liquid type, which is treated
in a fully gauge invariant Nambu-Gorkov formalism. Our work is thus a
generalization of the earlier work \cite{fertig,hwang} by Fertig and
Das Sarma, and by Hwang and Das Sarma (who considered
only the monolayer superconducting 
superlattice situation with only a single layer per unit cell) to a
complex unit cell situation with two layers per unit cell. To keep the
situation simple we will consider only the $s$-wave gap symmetry
\cite{fertig}, 
which, according to ref. \onlinecite{hwang} gives a very good account
of the collective 
mode dispersion even for the $d$-wave case except at very large wave
vectors. Following the work of Fertig and Das Sarma \cite{fertig}
there has been a 
great deal of theoretical and experimental work
\cite{hwang,cote,wu,for,das,kake,kado,dunm,buis} on the electronic
collective mode properties in layered superconducting materials, but
the specific issue considered in this paper has not earlier been
discussed in the 
literature for a multilayer superconducting system. It should also be
pointed out 
that, while the focus of our work is the collective mode behavior in
layered high-$T_c$ cuprate superconductors (which are {\it intrinsic}
superlattice systems due to their highly anisotropic crystal structure
with $CuO$ layers), our results equally well describe {\it artificial}
superconducting superlattices made of multilayer metallic structures
provided $k_{\|}$ and $k_{z}$ are good wave vectors in the system.

The collective mode dispersion in bilayered superconducting
superlattices is quite complicated.
There are essentially two different branches of long wavelength
collective modes: in-phase ($\omega_+$) modes and out-of-phase
($\omega_-$) modes, depending on whether the electron density
fluctuations in the two layers are in-phase or
out-of-phase. Each of these collective modes disperses as a function
of wave vector, showing strong anisotropy in $k_{\|}$ and $k_z$
dispersion. In particular, the limits ($k_z=0$, $k_{\|} \rightarrow
0$) and ($k_z \rightarrow 0 $, $k_{\|} = 0$) are {\it not} equivalent
because the $k_z=0$ three dimensional limit is singular.
For $k_z=0$ the in-phase $\omega_+$ collective mode is a gapped three
dimensional plasma mode at long wavelengths ($k_z=0$, $k_{\|}
\rightarrow 0$) by virtue of the Higgs mechanism arising from the long
range Coulomb interaction coupling all the layers. This mode
characterizes the long wavelength in-phase charge fluctuations of all
the layers. For non-zero $k_z$ the $\omega_+$ mode vanishes at long
wavelengths ($k_{\|} \rightarrow 0$) because at finite $k_z$ the
system is essentially two dimensional. The out-of-phase $\omega_-$
collective mode branch arises purely from the bilayer character of the
system and indicates the out-of-phase density fluctuations in the two
layers. In the absence of any interlayer hopping (either intracell and
intercell) the $\omega_-$ mode is purely acoustic in nature vanishing
at long wavelengths ($k_{\|} \rightarrow 0 $) as $\omega_-(k_z,k_{\|}
\rightarrow 0) \sim O(k_{\|})$ independent of the value of $k_z$. For
finite interlayer tunneling $\omega_-$ exhibits a tunneling gap at
$k_{\|} = 0$. The Higgs gap for $\omega_+(k_z=0, k_{\|} \rightarrow
0)$ is not qualitatively affected by intracell interlayer tunneling
because the three dimensional plasma energy is usually substantially
larger then the tunneling energy.

Note that, in the absence of any intracell and intercell tunneling,
both  in-phase and out-of-phase collective mode branches lie below
the superconducting energy gap for small $k_{\|}$ [except for the
$\omega_+(k_z=0)$ mode which is pushed up to the three dimensional
plasma frequency]. This remains true even for weak intracell and
intercell tunnelings, and in this paper we concentrate mainly on this
long wavelength ``below gap'' regime where the phase fluctuation modes
could possibly lie in the superconducting gap. For simplicity we also
restrict ourselves to $s$-wave gap symmetry of the superconducting
order parameter. This approximation may at first sight appear to be
unusually restrictive as it seems to rule out the applicability of our
theory to bilayer high-$T_c$ materials (such as YBCO, BISCO) which are
now widely accepted to have $d$-wave ground state symmetry. This,
however, is not the case because at long wavelengths (small $k_{\|}$),
which is what we mostly concentrate on, the collective mode spectrum is
insensitive to the order parameter symmetry \cite{hwang}, and
therefore our results 
apply equally well to high-$T_c$ bilayer materials. The modes we
predict and their dispersion should most easily be observable via the
resonant inelastic light scattering spectroscopy, but may also be
studied via frequency domain far infrared spectroscopy using a grating
coupler.

\section{theory, approximations, and results}

In our calculation we assume that the two layers in each unit cell can be 
considered to be 2D EG,  and all layers are identical, 
having the same 2D charge density $n_s$ per unit area.
Two identical layers separated by a
distance $c$ in each unit cell are strongly coupled through the
interlayer intracell 
electron tunneling. The interlayer tunneling is between the 
well-defined CuO layers in high T$_c$ materials. 
The intercell tunneling between different unit cells separated by a 
distance $d$ (in our model $d > c$) is
neglected at first (see section III for the effect of intercell
tunneling).  Although we neglect the electron tunneling between
different unit cells, the electrons in {\it all} layers are coupled
via the intercell long
range Coulomb potential which we keep in our theory.
Since the long wavelength plasma modes are independent of the gap
function symmetry 
\cite{hwang}, we work in the BCS approximation with s-wave pairing for
simplicity.  
Then, in the Nambu representation\cite{namb}
the effective Hamiltonian 
of a bilayered superconductor with 2D quasiparticle energy
$\varepsilon(k)$, a tight-binding coherent single-electron intracell
hopping $t(k)$, and an additional coherent intracell Josephson
coupling $T_J$ between two nearest layers is given by
\begin{equation}
H = H_0 - \mu N + H_{\rm int} + H_{T_J},
\label{ham}
\end{equation}
with
\begin{eqnarray}
H_0 - \mu N = & & \sum_{n,i} \sum_{\bf k} \tilde{\varepsilon}_{\bf k}
\Psi^{\dagger}_{{\bf k}, ni} \tau_3 \Psi_{{\bf k}, ni} \nonumber \\
&+& \sum_{n,i} \sum_{\bf k} t({\bf k})
\Psi^{\dagger}_{{\bf k}, ni} \tau_3 \Psi_{{\bf k}, n\bar{i}},
\label{h0}
\end{eqnarray}
\begin{equation}
H_{\rm int} = \frac{1}{2} \sum_{ni,mj} \sum_{\bf q}
\rho_{{\bf q},mi} \tilde{V}_{mi,nj}({\bf q}) \rho_{-{\bf q},nj} ,
\label{hint}
\end{equation}
\begin{equation}
H_{T_J} = \sum_{n,i}\sum_{{\bf k,k',q}}T_J \left ( \Psi_{{\bf k+q},ni}\tau_3 
\Psi_{{\bf k},n\bar{i}}\right ) \left ( \Psi_{{\bf k'-q},ni}\tau_3 
\Psi_{{\bf k'},n\bar{i}} \right ),
\label{htj}
\end{equation}
where $n$, $m$ are the unit cell indices and $i$, $j=1,2$ label the two
layers within a given unit cell ($\bar{i} = 3-i$).
Here, $\Psi_{{\bf k},ni}$ and $\Psi^{\dagger}_{{\bf k},ni}$ are the
column and row vectors
\begin{equation}
\Psi_{{\bf k},ni} \equiv \left ( \begin{array}{c}
c_{{\bf k},ni,\uparrow} \\ 
c^{\dagger}_{-{\bf k},ni,\downarrow}
\end{array} 
\right ), \;\;\;\;\;
\Psi^{\dagger}_{{\bf k},ni} \equiv \left (c^{\dagger}_{{\bf
k},ni,\uparrow}, c_{-{\bf k},ni,\downarrow} \right ), 
\label{psi}
\end{equation}
where $c^{\dagger}_{{\bf k},ni,\sigma}$ ($c_{{\bf k},ni,\sigma}$) creates
(annihilates) an
electron with wave vector ${\bf k}$ and spin 
$\sigma$ in the $i$-th layer within the $n$th unit cell,
and $\rho_{{\bf q},ni}$ denotes the density operator defined by
\begin{equation}
\rho_{{\bf q},ni} = \sum_{\bf k}\Psi^{\dagger}_{{\bf k+q},ni} \tau_3
\Psi_{{\bf k},ni},
\end{equation}
where
$\tilde {\varepsilon}_{\bf k} = 
{k^2}/{2m} -\mu$ ($\mu$ being the chemical potential), and $\tau_i$
($i$=1,2,3) are the Pauli matrices. In Eq. (\ref {hint}), the
effective interaction 
$\tilde{V}_{ni,mj}({\bf q})$ contains the long range Coulomb
interaction coupling all the layers and the short range attractive
intralayer pairing interaction (giving rise to superconductivity in
the problem)
\begin{equation}
\tilde{V}_{ni,mj}({\bf q}) = V_{c}(q_{\|}) \exp[-q_{\|}|z_{ni}-z_{mj}|]
+V_0 \delta_{n,m}\delta_{i,j}
\end{equation}
where $V_{c}(q_{\|}) = 2\pi e^2/(\kappa q_{\|})$ is the
two dimensional Coulomb interaction and $\kappa$ is the
background dielectric constant of the system.
$V_0$ represents a weak, short-ranged
attractive intra-layer pairing interaction which produces
superconductivity, and is a model parameter in our theory.

We should comment on one unusual feature of our Hamiltonian defined in
Eqs. (\ref{ham})--(\ref{htj}). This is the existence of {\it both} a
coherent single-particle 
hopping term, defined by the single-particle hopping amplitude $t(k)$
in Eq. (\ref{h0}), and a coherent Cooper pair Josephson tunneling
term, defined by $T_J$ in Eq. (\ref{htj}). Usually the existence of a
single-particle hopping $t$ 
automatically generates an equivalent Josephson coupling $T_J$ in the
superconducting system, and keeping both of them as we do, namely, $t$
in the single particle Hamiltonian $H_0$ [Eq. (\ref{h0})] and $T_J$ in the
two-particle Josephson coupling [Eq. (\ref{htj})], is redundant. We
do, however, 
wish to investigate separately effects of both coherent single
particle hopping and Josephson coupling along the $c$-axis on the
collective mode spectra because of recent suggestions \cite{anderson}
of a novel 
interlayer tunneling mechanism for superconductivity in cuprates which
explicitly postulates $t = 0$ (in the normal state) and $T_J \neq
0$ (in the superconducting state). Our model therefore uncritically
includes both $t$ and $T_J$ as distinct contributions, and one could
think of the interlayer Josephson coupling $T_J$ in our model
Hamiltonian arising from some interlayer pairing interaction not
included in our model pairing interaction $V_0$ which is exclusively
intralayer in nature. In the following we take $t$ and $T_J$ to be
independent parameters of our model without worrying about their
microscopic origins.

The collective modes of the system  are given by the poles of the
reducible density response function $\chi({\bf k},\omega)$. 
We apply the conserving gauge invariant ladder
diagram approximation\cite{fertig,namb} in calculating the density response
of the system including the effect of the pairing interaction induced
vertex and self-energy corrections. The density response function is
defined as 
\begin{equation}
\chi({\bf k},\omega)=-i\int^\infty_0 dt e^{i\omega t}  < \left [
\rho({\bf k},t), \rho(-{\bf k},0) \right ]  >,
\end{equation}
where $\rho({\bf k},t)$ is the three dimensional Fourier transform of
the density operator in the Heisenberg representation. Here,
${\bf k}\equiv(k_{\|},k_z) $ is the 3D wave
vector, where $k_z$ measures the wave vector along the $z$-axis (i.e., the
$c$-direction ) and $k_{\|}$ is the 2D {\it x-y} plane (i.e., {\it a-b} plane)
wave vector. The density response may be written 
in terms of an irreducible polarizability $\Pi({\bf k},\omega)$ as
shown in Fig. 1(a).
Following Anderson's arguments for the absence of 
single particle tunneling \cite{anderson} 
we first neglect inter-layer single 
electron
tunneling effects ($t=0$) and only consider the 

\begin{figure}
\epsfysize=6.0cm
\epsffile{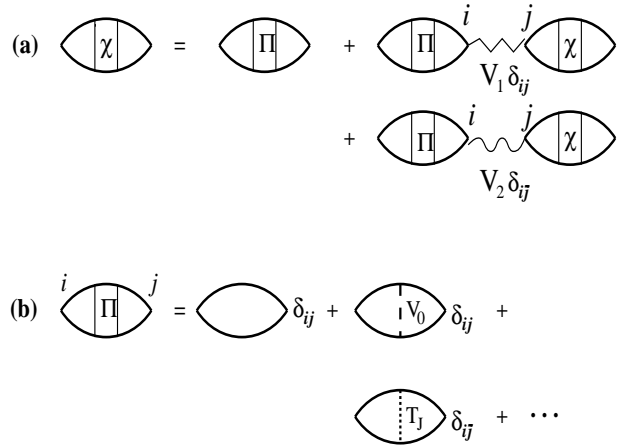}
\vspace{1.0cm}
\caption{
(a) Diagrammatic representation of the dielectric response in terms of the
irreducible polarizability $\Pi$. Here, $V_1$ and $V_2$ are given in
Eq. (\ref{v12}), and $\bar j = 3 - j$. 
(b) Irreducible polarizability used
in this calculation.}
\label{fig1}
\end{figure}

\noindent
Josephson coupling
effect. The polarizability $\Pi$ is then diagonal in the unit cell
index and becomes the corresponding 2D polarizability matrix,
$\Pi( {\bf k},\omega) \equiv \Pi(k_{\|},\omega)$
\begin{equation}
\chi({\bf k},\omega) = \frac{
\Pi({k_{\|}},\omega)}{\epsilon({\bf k},\omega)}, 
\label{chi}
\end{equation}
where $\Pi(k_{\|},\omega)$ is the irreducible polarizability matrix
for a single isolated unit cell,
\begin{equation}
\Pi({k_{\|}},\omega) = \left ( \begin{array}{cc}
\Pi_{11}({k_{\|}},\omega) & \Pi_{12}({k_{\|}},\omega) \\
\Pi_{21}({k_{\|}},\omega) & \Pi_{22}({k_{\|}},\omega)
\end{array} \right ),
\end{equation}
where $\Pi_{11}$, $\Pi_{22}$ and $\Pi_{12}$, $\Pi_{21}$ indicate the
intra-layer and inter-layer 
irreducible polarizability, respectively. 
Within our approximation, the inter-layer polarizabilities vanish when
the single-particle tunneling is neglected. We will see that the
plasma gap of the 
out-of-phase mode arises entirely from the non-vanishing inter-layer
irreducible polarizability. In Eq. (\ref{chi})
the effective dynamical dielectric function $\epsilon({\bf
k},\omega)$ is given by
\begin{equation}
\epsilon({\bf k},\omega) = {\bf 1} - \tilde{V}({\bf k})\Pi(k_{\|},\omega),
\label{eps}
\end{equation}
where {\bf 1} is a $2 \times 2$ unit matrix and $\tilde{V}({\bf k})$
is the effective 
interaction which in our simple model is given by
\begin{equation}
\tilde{V}(\bf k) = \left ( \begin{array}{cc}
V_{1}(\bf k) & V_{2}(\bf k) \\
V_{2}(\bf k) & V_{1}(\bf k)
\end{array} \right ),
\end{equation}  
where $V_1({\bf k})$ corresponds to the intralayer interaction ($i=j$)
and $V_2({\bf k})$ the interlayer interaction ($i\neq j$), which arises
entirely from the long-range Coulomb coupling in our model, and they
are given by 
\begin{eqnarray}
V_{1}({\bf k})& = & V_{c}(k_{\|}) f({\bf k})+ V_0, \nonumber \\ 
V_{2}({\bf k})& = & V_{c}(k_{\|}) g({\bf k}),
\label{v12}
\end{eqnarray}
where $f({\bf k})$ and $g({\bf k})$,
the superlattice form factors which modify the 2D Coulomb interaction
due to Coulomb coupling between all the layers in our multilayer
superlattice system, are given by
\begin{equation}
f({\bf k}) = \frac{\sinh (k_{\|}d)}{\cosh (k_{\|}d) - \cos (k_zd)},
\label{fq}
\end{equation}
\begin{equation}
g({\bf k}) = \frac{\sinh [k_{\|}(d-c)] + e^{-i k_z d} \sinh
(k_{\|}c)} {\cosh (k_{\|}d) - \cos (k_zd)}  e^{ik_zc}.
\label{gq}
\end{equation}

In order to obtain the collective mode spectrum, it is necessary to
construct a gauge invariant and number-conserving approximation for
$\Pi({\bf k},\omega)$. In the conserving gauge invariant ladder diagram
approximation\cite{fertig,namb} the irreducible polarizability 
obeys the ladder integral equation
which is shown diagrammatically in Fig. 1(b). 
It may be written in the form
\begin{eqnarray}
\Pi_{i,j}(k,\omega) = -i {\rm Tr}& &\int \frac{d\omega_1 dp_1}{(2\pi)^3}
\tau_3 G_{i}(p_1,\omega_1) \nonumber \\
& & \times \Gamma_{i,j}(p_1,k,\omega)
G_{i}(k-p_1,\omega-\omega_1), 
\end{eqnarray}
where $G_{i}(k,\omega)$ is the $i$-th layer Green's function with
self-energy corrections (self-consistent
Hartree approximation in 
the Coulomb interaction and self-consistent Hatree-Fock approximation
in the short-range pairing interaction)
and $\Gamma_{i,j}$  is a vertex function. 
The vertex part satisfies the linear integral equation 
\begin{eqnarray}
\Gamma_{ij}&(&p_1,k,\omega)  =  \tau_3 \delta_{ij}  +   i \sum_{l=1}^{2}
\int \frac{d^2q d\omega_1}{(2 \pi)^3} 
\tau_3 G_l(q,\omega_1) \nonumber \\
& \times & \Gamma_{ij}(q,k,\omega) 
G_l(q-k_1,\omega-\omega_1)\tau_3 \left [ V_0 \delta_{li} + T_J
\delta_{\bar{l} i} \right ],
\label{gaq1}
\end{eqnarray}
where $\bar{l} = 3-l$.
In order to solve this vertex function, we expand $\Gamma_{ij}$ in
Pauli matrices
\begin{equation}
\Gamma_{ij} = \sum_{l=0}^{3}\gamma_{ij,l}\tau_l.
\label{gaq2}
\end{equation} 
Since our model assumes two identical 2D layers in the unit cell, we have
$\Gamma_{11} = \Gamma_{22} = \Gamma_a$ and $\Gamma_{12} = \Gamma_{21}
= \Gamma_b$.
By introducing the polarization function
\begin{eqnarray}
P_{i} & = & i \int \frac{d^2qd\omega_1}{(2\pi)^3} \tau_3 G(q,\omega)
\tau_i G(q-k,\omega_1-\omega)\tau_3 \nonumber \\
        & = & \sum_{j=0}^{3} \bar{P}_{i,j}\tau_{j},
\label{pq}
\end{eqnarray}
the vertex function, Eq. (\ref{gaq1}), becomes
\begin{equation}
\left( \begin{array}{c}
\gamma_a \\ \gamma_b
\end{array} \right ) =
\left ( \begin{array}{c}
{\bf I}_3 \\ 0
\end{array} \right ) +
V_0 \left ( \begin{array}{c}
 \underline{P} \gamma_a \\ \underline{P} \gamma_b
\end{array} \right ) +
T_J \left ( \begin{array}{c}
\underline{P} \gamma_b \\ \underline{P} \gamma_a
\end{array} \right ),
\end{equation}
where $\gamma$'s are column vectors, $I_3^{T} = (0,0,0,1)$, and
$\underline{P}$ is a $4 \times 4$ 
matrix whose components are given by $\bar{P}_{ij}$ in Eq. (\ref{pq}). 
Then, the polarizability function $\Pi_{ij}$ becomes
\begin{eqnarray}
\Pi_{ij} & = & -{\rm Tr}\sum_{l=0}^{3} \bar{P}_{i,l}\tau_3 \gamma_{ij,l}
\nonumber \\ 
         & = & -\sum_{l=0}^{3} \left [ P_i \gamma_{ij} \right ]_{3,l}.
\label{e21}
\end{eqnarray}

The poles of the vertex function or polarizability $\Pi$  give the
collective mode spectra for the neutral superconductor (i.e.,
neglecting the long range Coulomb coupling which couples all the
layers). In 
the long wavelength limit we have two collective modes (``phasons'')
for the {\it neutral} system
\begin{equation}
\omega_{+}^2(k) = (v_0 k)^2 \left [ 1 + (V_0 + T_J)N_0/2 \right ],
\label{wpn}
\end{equation}
\begin{equation}
\omega_{-}^2(k) = \omega_0^2 + v_0^2 k^2
\left [ 1 + N_0 (V_0 - T_J)/2 \right ],
\label{wmn}
\end{equation}
where $v_0 = v_F/\sqrt{2}$ with $v_F$ as the Fermi velocity, $N_0 =
m/\pi$ is the 2D density of states at the Fermi surface, and $\omega_0^2 =
{16 T_J \Delta^2}/[N_0(V_0^2  - T_J^2)]$ is the tunneling gap induced
by the finite Josephson coupling 

\begin{figure}
\epsfysize=22.cm
\epsffile{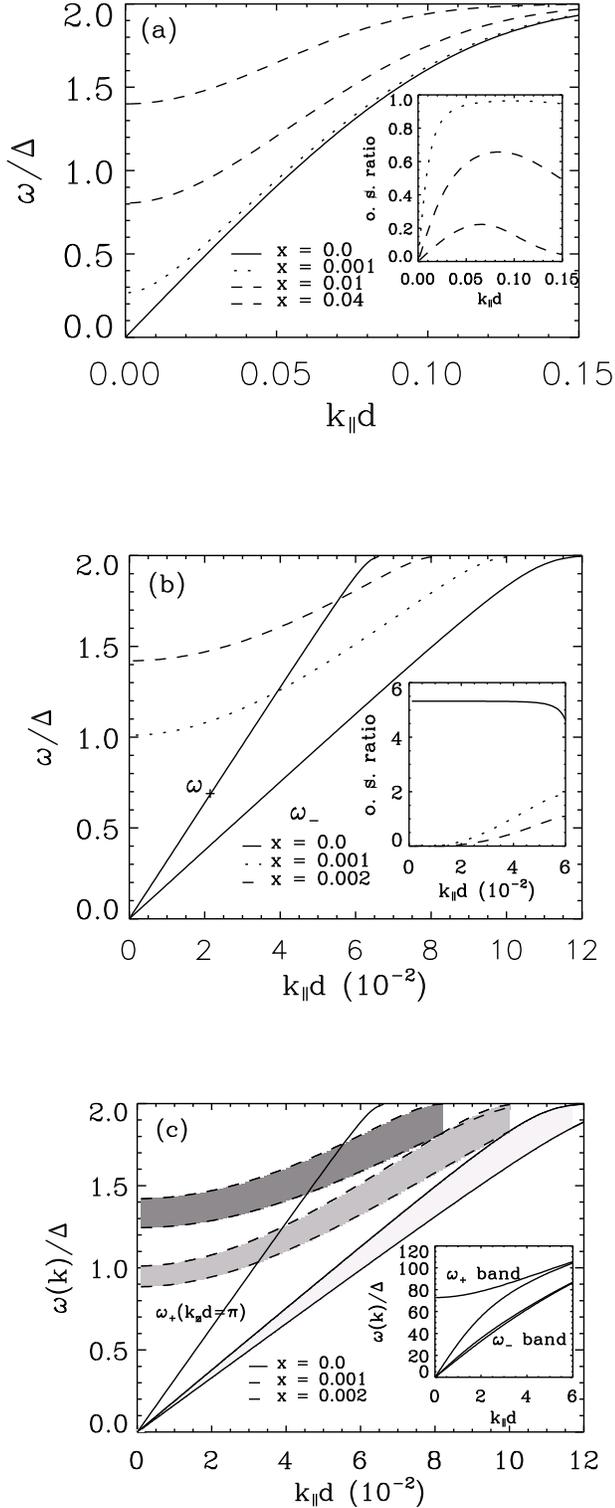}
\caption{
(a) The plasmon mode ($\omega_{\pm}$) dispersions in the presence of
Josephson 
tunneling for the neutral bilayered superconducting
superlattice as a function of $k_{\|}$ for fixed
$k_zd = \pi$.  Here, $x=T_J/V_0$ indicates the Josephson tunneling
strength with respect to the intra-layer pairing interaction. Inset
shows the ratio of the oscillator strength of $\omega_+$ to that of
$\omega_-$. 
(b) The plasmon mode dispersions ($\omega_{\pm}$) for the 
charged system. Inset shows the ratio of the oscillator strength of
$\omega_-$ to that of $\omega_+$.
(c) The $\omega_-({\bf k})$ band in the superlattice for the charged
system as a function 
of in-plane wave vector 
($k_{\|}d$) in the presence of the tunneling. Inset shows the
$\omega_{\pm}$ band of the bilayer superconducting superlattice. 
We use parameters roughly  
corresponding to YBCO in these figures: the sheet density $n_s =
10^{14} cm^{-2}$, 
effective in-plane mass $m=5m_0$, lattice dielectric constant $\kappa = 4$,
$d = 12 \AA$, and $c = 3\AA$.  
}
\label{fig2}

\end{figure}

\noindent
($T_J \neq 0$).
The $\omega_{+}$ mode corresponds to the in-phase motion of the order
parameter, or, equivalently the 2-D Goldstone-Anderson-Bogoliubov
phase fluctuation mode
due to the spontaneously broken continuous  gauge symmetry of the
superconducting state.   The $\omega_{-}$ mode corresponds to the
out-of-phase mode first predicted for a two-band superconductor
\cite{legg}, which has recently been 
calculated within the time-dependent Hartree-Fork-Gor'kov (mean-field)
approximation\cite{wu} for a two-layer superconductor system. In
Fig. \ref{fig2}(a) we show the calculated  
collective mode dispersion for different Josephson tunneling
strengths with respect to the intra-layer pairing interaction, $x =
T_J/V_0$. When the 
Josephson tunneling is absent, $x=0$, the two 
phason modes $\omega_{\pm}$ are degenerate and have identical
dispersion (solid line). 
But in the presence of finite Josephson tunneling between the nearest
layers, $x \neq 0$,
the out-of phase mode ($\omega_-$) develops a plasma gap ($\omega_0$)
depending on the 
tunneling strength.The in-phase mode 
$\omega_+$ is not affected qualitatively by finite Josephson tunneling
and remains an acoustic Goldstone mode (i.e., $\omega_+ \sim O(k)$ for $k
\rightarrow 0$) although the velocity of the acoustic plasmon does
depend on $T_J$ (cf. Eq. (\ref{wpn})).  In
Fig. \ref{fig2}(a) the inset shows the relative 
oscillator strength of the two phason modes, the ratio of the spectral
weight of $\omega_-$ to
that of $\omega_+$. The ratio  decreases as
tunneling amplitude increases. This is due to the approach of the
$\omega_-$ mode to the
pair-breaking excitation region ($\omega \approx 2\Delta$) at large
tunneling, which causes decay of the $\omega_-$ mode to single
particle excitations, and the
strength of the mode transfers to pair-breaking excitations. 
These results apply to any bilayered {\it neutral} superconductors (which,
of course, do not exist in nature because Coulomb interaction is
always present in real systems).

By looking for zeros of the dynamical dielectric function defined by
Eq. (\ref{eps}) we find the collective 
modes of the charged superconducting superlattices.
Since the two layers within the cell are identical we have
$\Pi_{11}=\Pi_{22}$ and $\Pi_{12} = \Pi_{21}$, which gives rise to
distinct in-phase and 
out-of-phase collective charge density fluctuations of the charged
superconductor. Coupling of the in-phase (out-of-phase) mode of the
neutral system via the long range Coulomb interaction
to the charge density fluctuation of the layers gives rise to the in-phase
(out-of-phase) collective mode of the charged bilayer system. The
dielectric function is a matrix, and 
the zeros of the det[$\epsilon$], which define the collective mode
spectra,  are  given by
\begin{eqnarray}
{\rm det}[\epsilon] = & & \left [ 1 - (\Pi_{11} + \Pi_{12})(V_1
+ V_2) \right ] \nonumber \\
& & \times \left [ 1- (\Pi_{11} - \Pi_{12})(V_1 - V_2) \right ] =0.
\label{e0}
\end{eqnarray}  
In the long wavelength limit Eq. (\ref{e0}) can be analytically
solved using Eqs. (\ref{v12}) -- (\ref{e21}), and we find two distinct
collective modes 
corresponding to the relative phase of the charge density fluctuations
in the two layers within each unit cell:
\begin{equation}
\omega_{+}^2({\bf k})  = \omega_p^2 \frac{k_{\|}d}{4} \left [
f({\bf k}) + |g({\bf k})| \right ]_{k_{\|}\rightarrow 0},
\label{wp}
\end{equation}
\begin{equation}
\omega_-^2({\bf k}) = \frac{ (1 + \Delta V - \Delta V_0 )(\omega_0^2 +
v_0^2k^2/2)}{1 - 
\omega_0^2(\Delta V - \Delta V_0 )/6},
\label{wmm}
\end{equation}
where $\omega_p=(4\pi n_Be^2/\kappa m)^{1/2}$ is a three dimensional
plasma frequency with the effective three-dimensional electron
density of the double-layered supperlattice $n_B = 2 n_s/d$,
and $k^2 = k_{\|}^2 + k_z^2$ with ${\bf k} \equiv (k_{\|}, k_z)$;
$\Delta V = N_0(V_1 - V_2)$ and $\Delta V_0 = N_0(V_0 - T_J)/2$. 
In Fig. \ref{fig2}(b) we show the calculated charge density mode
dispersion for  fixed $k_zd = \pi$ as a function of
$k_{\|}d$. Tunneling has little effect on the in-phase mode (thin
solid line)  but profoundly affects the out-of-phase mode (thick
lines) by introducing a gap at $\omega_-(k_{\|} = 0)$ similar to the
neutral case. Since in the
presence the tunneling the out-of-phase mode acquires a gap, the two
modes cross at the resonant frequency ($\omega_+ = \omega_-$), but the
symmetry (``parity'') associated with the two identical layers does not 
allow any mode coupling or anti-crossing effect. If the two layers in
the unit cell are  not identical then there is a mode coupling induced
anti-crossing around $\omega_+ \approx \omega_-$.
The inset shows the ratio of the oscillator strength of
the in-phase mode to that of the out-of-phase mode. In sharp contrast
to  the neutral system, in
the long wavelength limit  the
out-of-phase mode $\omega_-$ completely dominates the spectral weight
in the presence of interlayer tunneling.
In the absence of tunneling ($x = 0$), however, the in-phase mode
$\omega_+$ dominates the spectral weight.
Our results for the collective mode dispersion in the presence of
finite single-particle tunneling but vanishing Josephson coupling
(i.e., $t\neq 0$, $T_J =0$) are qualitatively identical to the
situation with $t=0$, $T_J \neq 0$, and are therefore not shown
separately. This is, of course, the expected result because $t$
automatically generates an effective Josephson tunneling, i.e., an
effective $T_J$, in the superconducting system, and therefore the
qualitative effect of having a finite $T_J$ or a finite $t$ in the
superconducting system is similar.

\begin{figure}
\epsfysize=7.1cm
\epsffile{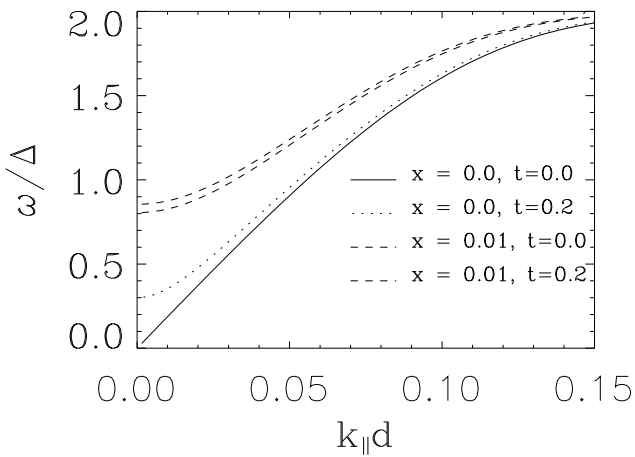}
\vspace{0.5cm}
\caption{The dispersion
of the out-phase mode ($\omega_-$) in the charged system in 
the presence of both the single particle 
tunneling and  Josephson tunneling as a function of $k_{\|}$ for a
fixed $k_z d = \pi$.
Here, $x = T_J/V_0$ and $t$ is the strength of the single particle
tunneling with respect to the superconducting energy gap. We
use the same parameters as Fig. \ref{fig2}.  
}
\label{fig3}
\end{figure}

We also calculate the collective modes of the bilayered superconducting
system by including {\it both} the
single particle tunneling and the Josephson tunneling
between the nearest layers (i.e., $t, T_J \neq 0$). The two layers in
the unit cell  
hybridized by the single particle tunneling matrix element, $t(\rm
k)$, would lead to a symmetric and an antisymmetric combination of the
quasiparticle states for each value of the wave vector ${\bf k}$ in
the plane.
By introducing the symmetric and antisymmetric single electron
operators with respect to 
an interchanging of the two layers, $\alpha_{n,k,\sigma} =
\frac{1}{\sqrt2}(c_{n1,k\sigma} + c_{n2,k\sigma})$ and
$\beta_{n,k,\sigma} = \frac{1}{\sqrt2}(c_{n1,k\sigma} - c_{n2,k\sigma})$, 
the total effective Hamiltonian can be written as
\begin{eqnarray}
H & & = 
\sum_n\sum_{k\sigma}\left [ \alpha_{n,k\sigma}^{\dagger}{\varepsilon_1}(k)
\alpha_{n,k\sigma} + \beta_{n,k\sigma}^{\dagger}{\varepsilon_2}(k)
\beta_{n,k\sigma} \right ] \nonumber \\
  &+ &  \frac{1}{2}\sum_{nn'}\sum_{\bf q} \left [
\rho_{1,n{\bf q}}^{T}\bar U({\bf q})\rho_{1,n'-{\bf q}} +
\rho_{2,n{\bf q}}^{T}\bar V({\bf q})\rho_{2,n'-{\bf q}} \right ],
\label{ham1}
\end{eqnarray}
where $\varepsilon_1(k) = \varepsilon(k) + t(k)$ and $\varepsilon_2(k)
= \varepsilon(k) - t(k)$,  
and 
\begin{eqnarray}
\rho_{1,n{\bf q}} & = & \sum_{{\bf k}\sigma}  
\left ( \begin{array}{c}
\alpha_{n,{\bf k}+{\bf q}\sigma}^{\dagger}\alpha_{n,{\bf k} \sigma} \\
\beta_{n,{\bf k}+{\bf q}\sigma}^{\dagger}\beta_{n,{\bf k} \sigma} 
\end{array} \right ),  \\
\rho_{2,n{\bf q}} & = & \sum_{{\bf k}\sigma} 
\left ( \begin{array}{c}
\alpha_{n,{\bf k}+{\bf q}\sigma}^{\dagger}\beta_{n,{\bf k} \sigma} \\
 \beta_{n,{\bf k}+{\bf q}\sigma}^{\dagger}\alpha_{n,{\bf k} \sigma}
\end{array} \right ),
\end{eqnarray}
and 
\begin{equation}
\bar U({\bf q}) = \left ( \begin{array}{cc}
U_+ & U_- \\
U_- & U_+
\end{array} \right ), \;\;\;\;\;
\bar V({\bf q}) = \left ( \begin{array}{cc}
V_+ & V_- \\
V_- & V_+
\end{array} \right ),
\end{equation}
where $U_{\pm} = V_1 + V_2 \pm T_J$ and $V_{\pm} = V_1 - V_2 \pm T_J$.
This Hamiltonian is identical to that in the corresponding two subband
model, which is well 
studied in semiconductor quantum well systems \cite{jain}. Since there
are no 
off-diagonal elements of the interaction with respect to the subband
index we have well separated intra-subband and inter-subband
collective modes corresponding to the 
in-phase and out-of-phase modes, respectively. Within our
gauge-invariant ladder diagram approximation  we can easily calculate
the mode dispersions by following the procedure outlined in
Eqs. (\ref{psi}) -- (\ref{wmm}).
The in-phase-mode for both
the neutral and the charged system is
insensitive to tunneling in the long wavelength limit, and has
essentially the same long 
wavelength dispersion as in 
Eq. (\ref{wpn}) and Eq. (\ref{wp}) respectively, up to second
order in $k$. The out-of-phase mode is, however, strongly affected by
both the coherent single 
particle tunneling and the Josephson tunneling, and has a dispersion
\begin{equation}
\omega_{-}^2(k) = \omega_0^2 + \left [ (2t)^2 +
v_0^2 k^2 \right ] 
\left [ 1 + \Delta V_0 \right ],
\label{wmc}
\end{equation}
for neutral superconductors, and
\begin{equation}
\omega_-^2({\bf k}) = \frac{\left ( 1 + \Delta V - \Delta V_0 \right )
\left [ \omega_0^2 + (2t)^2 + v_0^2k_{\|}^2 \right ]}{ 1 -
\frac{\omega_0^2}{6}\left ( \Delta V - \Delta V_0 \right )},
\end{equation}
for charged systems in the presence of finite tunneling.
In Fig. \ref{fig3}, we show the calculated mode dispersions as a
function of the in-plane wave vector $k_{\|}d$ for a fixed 

\noindent
$k_zd = \pi$. 
As emphasized before, the collective mode dispersion is qualitatively
independent of the specific tunneling mechanism (i.e., $t$ or $T_J$),
and therefore experiments involving collective modes cannot
distinguish between the existing tunneling mechanisms in high-$T_c$
superconductors as has recently been emphasized \cite{das} in a
related context.

\section{discussion and conclusion}

We calculate in this paper the collective charge density fluctuation
excitation spectra of both the
neutral and the charged superconducting bilayerd superlattices with
interlayer intra-cell single particle and Josephson tunneling. 
We use the conserving gauge-invariant ladder diagram
approximation in the Nambu-Gorkov  formalism. In general, there are
two types of density fluctuation modes: in-phase ($\omega_+$) and
out-of-phase ($\omega_-$) modes.  
For neutral superconductors, the out-of-phase collective mode with
interlayer tunneling has a
plasma gap depending on the tunneling intensity, and the in-phase
mode, lying lower in energy, dominates
the oscillator strength for all wave vectors. However, for
charged superconductors the two phase modes couple to the long range
Coulomb interaction differently, and
the out-of-phase mode with tunneling dominates the oscillator strength
in the long wavelength limit ($k_{\|} \rightarrow 0$) and finite $k_z$.
Since we have used two identical 2D layers in each unit cell there is
no mode coupling effect in our theory between $\omega_{\pm}$ modes at
the resonant 
frequency ($\omega_+ \sim \omega_-$).
If the two layers forming the unit cell are not identical then there
will be a resonant mode coupling effect (``anti-crossing'') between
the in-phase and the out-of-phase modes around $\omega_+ \approx
\omega_-$ resonance point -- the nature of this anti-crossing
phenomena will be similar to what is seen in the corresponding
intrasubband-intersubband collective mode coupling in semiconductor
quantum well systems \cite{jain}. We have mostly concentrated in the long
wavelength regime ($k_{\|} \rightarrow 0$) -- at large wave vectors
there is significant 
coupling between the collective modes and the pair-breaking
excitations, which has been extensively studied in the literature
\cite{fertig,hwang}. We 
have also neglected the amplitude fluctuation modes because they
usually carry negligible spectral weights compared with the
$\omega_{\pm}$ phase fluctuation modes. We have also used an $s$-wave
ground state symmetry which should be a good approximation
\cite{hwang} even for $d$-wave cuprate systems as far as the long
wavelength collective mode properties are concerned. Our use of a
BCS--Fermi liquid model in our theory is more difficult to defend
except on empirical grounds \cite{das} and for reasons of simplicity.

Finally, we consider the effect of {\it intercell} tunneling on the
collective mode spectra, which we have so far neglected in our
consideration. (Our theory includes both intracell and intercell
Coulomb coupling between all the layers, and {\it intracell}
interlayer single electron and Josephson tunneling.) The neglect of
intercell tunneling is justified by the fact that $d \gg c$ (e.g., in
YBCO $d = 12 \AA$, $c=3 \AA$). The general
effect of intercell tunneling becomes quite complicated theoretically
because one has far too many interlayer coupling terms in the
Hamiltonian in the presence of {\it both} intracell and 
intercell interlayer tunneling involving {\it both} single particle and
Josephson tunneling. It is clear, however, that the main effect of a
weak intercell interlayer tunneling (either single particle or
Josephson type, or both) would be to cause a 2D to 3D transition in
the plasma mode by opening up a small gap in both $\omega_{\pm}$ modes
at long wavelengths (in the charged system). The size of this gap
(which is the effective 3D plasma frequency of the $k_z$-motion
of the system) will depend on the intercell tunneling strength.
This small gap is the 3D $c$-axis plasma frequency of the system,
which has been the subject of several recent studies in the literature
\cite{das,anderson,tama}.

The introduction of a weak {\it intercell} interlayer tunneling will
therefore modify our calculated results simply through a shift of the
energy/frequency origin in our calculated dispersion curves. The
origin of the ordinate (i.e., the energy/frequency axis) in our
results will shift from zero to $\omega_c$, where $\omega_c$ is the
c-axis plasma frequency arising from the {\it intercell} interlayer
hopping. For an effective single band tight binding intercell hopping
parameter $t_c$ (i.e., the single electron effective bandwidth in the
$c$-direction is $2t_c$), one obtains $\omega_c = \omega_p t_c d/v_F$,
where $\omega_p=[4\pi n_B e^2/(\kappa m)]^{1/2}$ is the effective 3D
plasma frequency with the 2D {\it a-b} plane band mass $m$ [see
Eq. (\ref{wp})] and $v_F$ is the Fermi velocity in the {\it a-b}
plane. Note that $\omega_c \ll \omega_p$ because $t_c$ is very small
by virtue of weak intercell coupling. Note also that if one defines an
effective ``3D'' $c$-axis plasma frequency $\omega_{pc}=[4\pi n_B
e^2/(\kappa m_c)]^{1/2}$ in analogy with $\omega_p$, where $m_c$ is
now the effective mass for electron dynamics along the $c$-axis, then
$\omega_c = \omega_{pc} [t/(2E_F)]^{1/2}$ due to the tight bind nature
of $c$-motion. We emphasize that in the presence of intercell hopping
$\omega_c$   sets the scale for the lowest energy that a collective
mode can have in the multilayer superconductor -- $\omega_c$ is
sometimes referred \cite{cote,kake,kado} to as a Josephson plasmon
\cite{anderson} in the literature. In general, it is difficult to
theoretically estimate $\omega_c$ in high-$T_c$ materials \cite{das}
because the effective $t_c$ (and other parameters) may not be known. It
is therefore important to emphasize \cite{das,anderson} that
$\omega_c$ can be measured directly from the $c$-axis plasma edge in
reflectivity experiments, (we emphasize that {\it a-b} plane plasma
edge gives $\omega_p$ and the $c$-axis plasma edge gives $\omega_c$
\cite{tama}), and such measurements \cite{tama} show that $\omega_c$
is below the superconducting gap in many high-$T_c$
materials \cite{das}. 
This implies that the effective $c$-axis hopping, $t_c$, in high-$T_c$
materials (either due to single particle hopping or due to Josephson
coupling arising from coherent Cooper pair hopping) has to be very
small (much smaller than that given by direct band structure
calculations) in these systems for the Josephson plasma frequency
$\omega_c$ to be below the superconducting gap, a point first
emphasized by Anderson \cite{anderson}.

The collective mode situation in a bilayer system in the presence of
both intracell and intercell interlayer coupling is obviously quite
complex, and as emphasized in ref. \onlinecite{anderson}, there could
in general be several collective phase fluctuation modes depending on
the detailed nature of intracell and intercell interlayer hopping
matrix. 
In the most general bilayer system intercell coupling will give rise
to two separate $\omega_+$ plasma bands arising from the two distinct
possible intercell interlayer coupling --- the two $\omega_+$ bands
lying in energy lower that the two $\omega-$ bands in the charged
system as we show in this paper. 
In the most general situation \cite{anderson}, there could be
two low energy Josephson plasma frequencies $\omega_{c1}$,
$\omega_{c2}$ ($>$$\omega_{c1}$), corresponding to the bottoms of the
two $\omega_+$ bands, arising respectively from the larger
and the smaller of the intercell interlayer hopping amplitudes. To
make things really complicated one of these modes ($\omega_{c1}$)
could be below the gap and the other ($\omega_{c2}$) above the
gap, (or, both could be below or above the gap). While each of these
scenarios  is  possible, $c$-axis optical response
experimental results on inter-bilayer charge dynamics in $YBCO$ have
been interpreted \cite{cooper} to exhibit only one $c$-axis plasma
edge in the superconducting state with the frequency $\omega_c$
between 60 cm$^{-1}$ and 200 cm$^{-1}$, depending on the oxygen
content. There are three possibilities: (1) The two plasma modes
($\omega_{c1} \approx \omega_{c2} \approx \omega_c$) are almost
degenerate because the corresponding intercell hopping amplitudes are
close in magnitudes; (2) $\omega_{c2}$ is much lager than $\omega_{c1}$
($\ll \omega_{c2}$) because the two intercell hopping amplitude are
very different in magnitudes (we consider this to be an unlikely
scenario); (3) one of the two modes carries very little optical
spectral weight and is not showing up in $c$-axis reflectivity
measurements, leaving only the other one as the observed $c$-axis
plasma edge. There is, in principle, a fourth (very unlikely)
possibility: the observed plasma edge is really $\omega_{c2}$, and the
other mode $\omega_{c1}$ ($\ll \omega_{c2}$) is too low in energy to
show up in $c$-axis reflectivity measurements.

Within a {\it nearest-neighbor} $c$-axis interlayer coupling model, there is
only a {\it single} intercell hopping amplitude, giving rise to only a
single $c$-axis plasma edge $\omega_c$, which now defines the lowest
value that the in-phase collective mode $\omega_+$ can have, $\omega_c
\equiv \omega_{c+} \equiv \omega_+(k=0)$ --- $\omega_c$ is shifted up
from zero at long wavelengths due to finite $c$-axis intercell
hopping. The out-of-phase plasma edge, $\omega_{c-} \equiv
\omega_{-}(k=0)$, will obviously lie much higher in energy than
$\omega_{c+} \equiv \omega_c$ because the intracell interlayer hopping
is much stronger than the intercell interlayer hopping. In particular,
even though the $\omega_{c+}$ mode may lie in the superconducting gap
\cite{cooper,anderson}, we expect $\omega_{c-}$ to
lie much above the superconducting gap energy in $YBCO$. A crude
qualitative estimate can be made by assuming that the intra- and
intercell hopping amplitudes scale as inverse squares of lattice
parameters: $t_{\rm intra}/t_{\rm inter} \approx (d/c)^2 = 16$. This
then leads to the approximate formula $\omega_{c-} \approx 16^2 \;
\omega_{c+} = 256 \; \omega_c$, which, for $YBCO$, implies that the
long wavelength out-of-phase mode should lie between 2 eV and 6 eV,
depeding on the oxygen content (assuming that the c-axis plasma edge
varies between 60 cm$^{-1}$ and 200 cm$^{-1}$, as reported in
ref. \onlinecite{cooper}, depending on the oxygen content). While
there is some minor observable structure in optical experiments at
high energies, we cannot find any compelling evidence in favor of the
existence of a high energy out-of-phase mode in the currently
available experimental data. We feel that a spectroscopic experiment,
using, for example, the inelastic electron energy loss spectroscopy
which could probe the mode dispersion (and which has been highly
successful in studying bulk plasmons in metal films) of the $\omega_-$
mode at high energy, may be required to unambiguously observe the
out-of-phase collective mode.
What we have shown in this paper is that under suitable conditions
(finite $k$ and $k_z$) the $\omega_-$ out-of-phase mode carries
reasonable spectral weight and should be observable in principle ---
actual observation, however, awaits experimental investigations using
external probes which can study mode dispersion at finite wave vectors
(which optical experiments by definition cannot do; they are long
wavelength probes).

\section*{ACKNOWLEDGMENTS}

This work is supported by the U.S.-ARO and the U.S.-ONR.

\end{document}